# On Pre-transformed Polar Codes

Bin Li, Huazi Zhang, Jiaqi Gu
Dept. Communication Technology Research, Huawei Technologies, P. R. China
{binli, zhanghuazi, gujiaqi}@huawei.com

*Abstract*— In this paper, we prove that any pre-transformation with an upper-triangular matrix (including cyclic redundancy check (CRC), parity-check (PC) and convolution code (CC) matrix) does not reduce the code minimum distance and an properly designed pre-transformation can reduce the number of codewords with the minimum distance. This explains that the pre-transformed polar codes can perform better than the Polar/RM codes.

## I. INTRODUCTION

Polar codes are a major breakthrough in coding theory [1]. They can achieve Shannon capacity with a simple encoder and a simple successive cancellation decoder when the code block size is large enough. But for moderate lengths, the error rate performance of polar codes with the SC decoding is not as good as LDPC or turbo codes. A SC-list decoding algorithm was proposed for polar codes [2], which performs better than the simple SC decoder and performs almost the same as the optimal ML (maximum likelihood) decoding at high SNR. In order to improve the minimum distance of polar codes, either RM-Polar codes [3], or the concatenation of polar codes with CRC [2][4] and PC [5] were proposed to significantly enhance error-rate performance. Recently, a new PAC (polarization-adjusted convolutional) code was proposed [6], by performing a convolution operation before RM (128,64) code, the PAC (128,64) code can provide a much better error-rate performance than RM (128,64) code. In this paper, we show that pre-transformed polar codes have better distance spectrum than Polar/RM codes in terms of the minimum distance and the number of codewords with the minimum distance, which can explain why PAC (128,64) can perform much better than RM (128,64) under ML-type decoding.

In section II, we review the encoding of Polar, RM and pre-transformed polar codes and in section III we prove that any pre-transformation with an upper-triangular matrix (including convolution matrix) does not reduce the code minimum distance. In Section IV, we prove that the pre-transformation with an upper-triangular matrix (including convolution matrix) can reduce the number of codewords with the minimum distance. Finally we draw some conclusions in section V.

## II. POLAR, RM AND PAC CODES

### A. Encoding of Polar Codes

$F = \begin{bmatrix} 1 & 0 \\ 1 & 1 \end{bmatrix}$, $F^{\otimes n}$ is a $N \times N$ matrix, where $N = 2^n$, $\otimes n$ denotes $n$th Kronecker power, and $F^{\otimes n} = F \otimes F^{\otimes(n-1)}$. Let $H_N = F^{\otimes n}$, the Polar/RM codes can be generated as

$$X = U \times H_N \quad (1)$$

where $U = (u_1, u_2, \ldots, u_N)$ is the encoded bit sequence. According to the principle of Polar design, these encoding bits $(u_1, u_2, \ldots, u_N)$ have different reliabilities, and these $N$ bits are divided into two subsets according to their reliabilities. The top $K$ most reliable bits are used to send information and the rest are frozen bits set to zeros. But for RM codes, the K bits are selected according to the weights of their corresponding rows in matrix $H_N$. The N encoding bits have different weights, the K bits with the top largest weights are selected as information bits and the rest of bits are frozen bits set to zeros.

### B. Pre-transformed Polar Codes

The pre-transformed polar codes are generated as

$$X = U \times T \times H_N \quad (2)$$

where T is an upper-triangular matrix with elements: $T_{i,j} = 0$, if $i > j$; $T_{i,j} = 1$, if $i = j$; $T_{i,j} \in \{0,1\}$, if $i < j$. The minimum distance $d_{min}$ is defines as the minimum weight among all non-zero codewords, and let the number of the codewords with the minimum distance be $N_{min}$.

It is easy to verify that outer concatenation of polar codes with CRC [2][3] and PC [5], and the recently proposed PAC [6], fall into the category of pre-transformed polar codes.

## III. THE MINIMUM DISTANCE: $d_{min}(U \times T \times H_N) \geq d_{min}(U \times H_N)$

Let $w(X)$ be the weight of $X$, $h_N^{(m)}$ be the $m$th row vector of $H_N$, and $w(h_N^{(m)})$ be the weight of $h_N^{(m)}$, $1 \leq m \leq N$.

*Corollary 1*: For any $c_{m+1}, c_{m+2}, \ldots, c_N \in \{0,1\}$, $w\left(h_N^{(m)} \oplus \left(\sum_{k=m+1}^{N} c_k h_N^{(k)}\right)\right) \geq w(h_N^{(m)})$, where $1 \leq m \leq N$, $\oplus$ is the XOR operation between row vectors.

*Proof:* 1) when N=4, it is easy to check that $w\left(h_4^{(m)} \oplus \left(\sum_{k=m+1}^{4} c_k h_4^{(k)}\right)\right) \geq w(h_4^{(m)})$, for $1 \leq m \leq 4$.

2) Induction hypothesis that $w\left(h_N^{(m)}\oplus(\sum_{k=m+1}^{N}c_kh_N^{(k)})\right) \geq w(h_N^{(m)})$, where $1 \leq m \leq N$, we need to prove that $w\left(h_{2N}^{(m)}\oplus(\sum_{k=m+1}^{2N}c_kh_{2N}^{(k)})\right) \geq w(h_{2N}^{(m)})$, where $1 \leq m \leq 2N$.

Case I: $1 \leq m \leq N$
If $1 \leq k \leq N$, $h_{2N}^{(k)} = [h_N^{(k)} \ Z_N]$, where $Z_N = (0\ 0\ ...\ 0)$ which has a length of N, then $w(h_{2N}^{(k)}) = w(h_N^{(k)})$; If $N+1 \leq k \leq 2N$, $h_{2N}^{(k)} = [h_N^{(k-N)} \ h_N^{(k-N)}]$, then $w(h_{2N}^{(k)}) = 2 \times w(h_N^{(k-N)})$.

$$h_{2N}^{(m)}\oplus(\sum_{k=m+1}^{2N}c_kh_{2N}^{(k)})$$
$$= h_{2N}^{(m)}\oplus(\sum_{i=m+1}^{N}c_ih_{2N}^{(i)})\oplus\left(\sum_{j=N+1}^{2N}c_jh_{2N}^{(j)}\right) \quad (3)$$
$$= [h_N^{(m)}\ Z_N]\oplus[(\sum_{i=m+1}^{N}c_ih_N^{(i)})\ Z_N]$$
$$\oplus\left[\left(\sum_{j=1}^{N}c_{j+N}h_N^{(j)}\right)\ \left(\sum_{j=1}^{N}c_{j+N}h_N^{(j)}\right)\right] \quad (4)$$
$$= [h_N^{(m)}\oplus(\sum_{i=m+1}^{N}c_ih_N^{(i)})\ Z_N]$$
$$\oplus\left[\left(\sum_{j=1}^{N}c_{j+N}h_N^{(j)}\right)\ \left(\sum_{j=1}^{N}c_{j+N}h_N^{(j)}\right)\right] \quad (5)$$
$$= \left[h_N^{(m)}\oplus(\sum_{i=m+1}^{N}c_ih_N^{(i)})\oplus\left(\sum_{j=1}^{N}c_{j+N}h_N^{(j)}\right)\ \left(\sum_{j=1}^{N}c_{j+N}h_N^{(j)}\right)\right] \quad (6)$$

$$w\left(h_{2N}^{(m)}\oplus(\sum_{k=m+1}^{2N}c_kh_{2N}^{(k)})\right)$$
$$= w\left(h_N^{(m)}\oplus(\sum_{i=m+1}^{N}c_ih_N^{(i)})\oplus\left(\sum_{j=1}^{N}c_{j+N}h_N^{(j)}\right)\right) +$$
$$w\left(\sum_{j=1}^{N}c_{j+N}h_N^{(j)}\right) \quad (7)$$
$$\geq$$
$$w\left(h_N^{(m)}\oplus(\sum_{i=m+1}^{N}c_ih_N^{(i)})\oplus\left(\sum_{j=1}^{N}c_{j+N}h_N^{(j)}\right)\oplus\left(\sum_{j=1}^{N}c_{j+N}h_N^{(j)}\right)\right)$$
$$= w\left(h_N^{(m)}\oplus(\sum_{i=m+1}^{N}c_ih_N^{(i)})\right) \quad (8)$$

According to the induction hypothesis that
$w\left(h_N^{(m)}\oplus(\sum_{k=m+1}^{N}c_kh_N^{(k)})\right) \geq w(h_N^{(m)})$
$$w\left(h_{2N}^{(m)}\oplus(\sum_{k=m+1}^{2N}c_kh_{2N}^{(k)})\right) \geq w(h_N^{(m)}) = w(h_{2N}^{(m)}) \quad (9)$$

Case II: $N+1 \leq m \leq 2N$
$$h_{2N}^{(m)}\oplus(\sum_{k=m+1}^{2N}c_kh_{2N}^{(k)}) = [h_N^{(m-N)}\ h_N^{(m-N)}]$$
$$\oplus[(\sum_{k=m-N+1}^{N}c_{k+N}h_N^{(k)})\ (\sum_{k=m-N+1}^{N}c_{k+N}h_N^{(k)})] \quad (10)$$

$$= [h_N^{(m-N)}\oplus(\sum_{k=m-N+1}^{N}c_{k+N}h_N^{(k)})$$
$$h_N^{(m-N)}\oplus(\sum_{k=m-N+1}^{N}c_{k+N}h_N^{(k)})] \quad (11)$$

$$w\left(h_{2N}^{(m)}\oplus(\sum_{k=m+1}^{2N}c_kh_{2N}^{(k)})\right)$$
$$= 2 \times w\left(h_N^{(m-N)}\oplus(\sum_{k=m-N+1}^{N}c_{k+N}h_N^{(k)})\right)$$
$$\geq 2 \times w(h_N^{(m-N)}) = w(h_{2N}^{(m)}) \quad (12)$$

**Theorem 1**: For any T with $T_{i,j} = 0$, if $i > j$; $T_{i,j} = 1$, if $i = j$; $T_{i,j} \in \{0,1\}$, if $i < j$. The minimum distance $d_{min}(U \times T \times H_N) \geq d_{min}(U \times H_N)$.

*Proof:* For the $U \times H_N$ code, the minimum distance is $d_{min}(U \times H_N) = \min_{m \in bit\ index}\left(w(h_N^{(m)})\right)$. For any non-all-zero sequence U, without loss of generality, let $m$ be the first non-zero information bit index, i.e., $U = (0\ 0\ ...,0, u_m = 1, u_{m+1},...,u_N)$, this is because that the frozen bits are zero, the first non-zero bit must be an information bit, then we have
$$U \times T = (0\ 0\ ...,0, v_m = 1, v_{m+1},...,v_N) \quad (12)$$
$$U \times T \times H_N = h_N^{(m)}\oplus(\sum_{k=m+1}^{N}v_kh_N^{(k)}) \quad (13)$$
$$w(U \times T \times H_N) = w\left(h_N^{(m)}\oplus(\sum_{k=m+1}^{N}v_kh_N^{(k)})\right) \geq w(h_N^{(m)})$$
$$\geq \min_{m \in bit\ index}\left(w(h_N^{(m)})\right) = d_{min}(U \times H_N) \quad (14)$$
Since for any non-all-zero codeword U, the weight $w(U \times T \times H_N) \geq d_{min}(U \times H_N)$, therefore we have $d_{min}(U \times T \times H_N) \geq d_{min}(U \times H_N)$.

## IV. THE NUMBER OF CODEWORDS WITH THE MINIMUM DISTANCE $N_{min}(U \times T \times H_N) \leq N_{min}(U \times H_N)$

**Theorem 2**: If $U \times H_N$ have the minimum distance $d_{min}$, and the second least minimum distance larger than $d_{min} + 2$, then there is a T such that $N_{min}(U \times T \times H_N) \leq N_{min}(U \times H_N)$, where T is an upper-triangular matrix with elements: $T_{i,j} = 0$, if $i > j$; $T_{i,j} = 1$, if $i = j$; $T_{i,j} \in \{0,1\}$, if $i < j$.

*Proof:* Suppose that $U \times H_N$ contains $N_{min}$ codewords: $D_k = (d_{k,1}, d_{k,2},...,d_{k,N})$, where $1 \leq k \leq N_{min}$, which have the minimum distance $d_{min}$. Let $\{I_1, I_2, ..., I_q\}$ be the information bit indices which is equal to or less than or N/2, i.e., $I_m \leq N/2$, $1 \leq m \leq q$. We will design T by computing the impact of $h_N^{(I_m)}\oplus h_N^{(N/2+1)}$ on $N_{min}(U \times H_N)$ as follows:

$h_N^{(N/2+1)}$ contains two 1's at position 1 and N/2+1, respectively. Let us consider the bit patterns of two bits $(d_{k,1}, d_{k,N/2+1})$ in $D_k$, $1 \leq k \leq N_{min}$. Let $w_j$ ($1 \leq j \leq q$) be the number of patterns of $(d_{k,1} = 0, d_{k,N/2+1} = 0)$ corresponding to the information bit at index $I_j$ is "1", i.e., $u_{I_j} = 1$. If T is constructed as follows: $T_{i,i} = 1$ ($1 \leq i \leq N$), $T_{I_j,N/2+1} = 1$, and other else are zeros, then $U \times T \times H_N$ contains $N_{min} - w_j$ codewords with the minimum distance $d_{min}$. The reason is as follows:
Case 1. $U \times H_N \in D_k$:
The set $D_k$ is divided into two subsets according to $u_{I_j} = 0$ and $u_{I_j} = 1$. When $u_{I_j} = 0$, $U \times H_N$ and $U \times T \times H_N$ generate the same codeword, i.e., $U \times T \times H_N = U \times H_N$. When $u_{I_j} = 1$, $U \times T \times H_N = (U \times H_N) \oplus h_N^{(N/2+1)}$. If $U \times H_N$ contains pattern $(d_{k,1} = 0, d_{k,N/2+1} = 0)$, $U \times T \times H_N$ will have a weight of $d_{min} + 2$; if $U \times H_N$ contains pattern $(d_{k,1} = 0, d_{k,N/2+1} = 1)$ or $(d_{k,1} = 1, d_{k,N/2+1} = 0)$, $U \times T \times H_N$ will have a same weight of $d_{min}$. Fortunately, $U \times H_N$ does not contain the bit pattern $(d_{k,1} = 1, d_{k,N/2+1} = 1)$ which leads to that $U \times T \times H_N$ have a weight $d_{min} - 2$, this is because that $U \times T \times H_N$ does not reduce $d_{min}$;

Case 2: $U \times H_N \notin D_k$
When $U \times H_N \notin D_k$, $w(U \times H_N) > d_{min} + 2$. $w(U \times T \times H_N) = w\left((U \times H_N) \oplus h_N^{(N/2+1)}\right) > d_{min}$.

Let $j_{opt} \underset{j}{\rightarrow} max(w_j)$, $T_{i,i} = 1$ ($1 \leq i \leq N$) and $T_{I_{j_{opt}},N/2+1} = 1$, then $U \times T \times H_N$ will have the least number of codewords with the minimum distance, which is $N_{min} - max(w_j)$.

For example, RM(32,16) code with information bit indices {8,12,14,15,16,20,22,23,24,26,27,28,29,30,31,32} have $d_{min} = 8$, $N_{min} = 620$, and the second least distance is 12. q=5, $\{I_1, I_2, I_3, I_4, I_5\} = \{8,12,14,15,16\}$. TABLE I shows the $w_j$ values, where $1 \leq j \leq 5$. TABLE II shows the number of codewords of $d_{min}$ for different $T$ designed by $h_N^{(I_m)} \oplus h_N^{(N/2+1)}$, where $1 \leq m \leq 5$. It is shown that $T$ reduces the number of codewords with weight of 8 from 620 to 492.

TABLE I. $w_j \sim h_N^{(I_j)} \oplus h_N^{(N/2+1)}$ ($1 \leq j \leq q$) of RM(32,16)

| $w_1$ | $w_2$ | $w_3$ | $w_4$ | $w_5$ |
|---|---|---|---|---|
| 128 | 128 | 128 | 128 | 128 |

TABLE II. $N_{min}(U \times T \times H_N)$ for Different $T$s.

| $T_{i,i} = 1$ ($1 \leq i \leq 32$) | $T_{i,i} = 1$ ($1 \leq i \leq 32$) | $T_{i,i} = 1$ ($1 \leq i \leq 32$) | $T_{i,i} = 1$ ($1 \leq i \leq 32$) | $T_{i,i} = 1$ ($1 \leq i \leq 32$) |
|---|---|---|---|---|
| $T_{8,17} = 1$ | $T_{12,17} = 1$ | $T_{14,17} = 1$ | $T_{15,17} = 1$ | $T_{16,17} = 1$ |
| 492 | 492 | 492 | 492 | 492 |

TABLE III. $w_j \sim h_N^{(I_j)} \oplus h_N^{(N/2+1)} \oplus h_N^{(N/2+2)}$ ($1 \leq j \leq q$).

| $w_1$ | $w_2$ | $w_3$ | $w_4$ | $w_5$ |
|---|---|---|---|---|
| 128 | 128 | 128 | 128 | 128 |

TABLE IV. $N_{min}(U \times T \times H_N)$ for Different $T$s.

| $T_{i,i} = 1$ ($1 \leq i \leq 32$) | $T_{i,i} = 1$ ($1 \leq i \leq 32$) | $T_{i,i} = 1$ ($1 \leq i \leq 32$) | $T_{i,i} = 1$ ($1 \leq i \leq 32$) | $T_{i,i} = 1$ ($1 \leq i \leq 32$) |
|---|---|---|---|---|
| $T_{8,17} = 1$ | $T_{12,17} = 1$ | $T_{14,17} = 1$ | $T_{15,17} = 1$ | $T_{16,17} = 1$ |
| $T_{8,18} = 1$ | $T_{12,18} = 1$ | $T_{14,18} = 1$ | $T_{15,18} = 1$ | $T_{16,18} = 1$ |
| 492 | 492 | 492 | 492 | 492 |

We also can design $T$ by computing the impact of $h_N^{(I_m)} \oplus h_N^{(N/2+1)} \oplus h_N^{(N/2+2)}$ on $N_{min}(U \times H_N)$. $h_N^{(N/2+1)} \oplus h_N^{(N/2+2)}$ contains two 1's at position 2 and N/2+2, respectively. Let us consider the bit patterns of two bits $(d_{k,2}, d_{k,N/2+2})$ in $D_k$, $1 \leq k \leq N_{min}$. Considering the bit patterns of two bits $(d_{k,2}, d_{k,N/2+2})$ in $D_k$, $1 \leq k \leq N_{min}$. Let $w_j$ be the number of patterns of $(d_{k,2} = 0, d_{k,N/2+2} = 0)$ corresponding to the information bit at index $I_j$ is "1", i.e., $u_{I_j} = 1$. If $T_{i,i} = 1$ ($1 \leq i \leq N$), $T_{I_j,N/2+1} = 1$, $T_{I_j,N/2+2} = 1$, and other else are zeros, then $U \times T \times H_N$ contains $N_{min} - w_j$ codewords with the minimum distance $d_{min}$.

For example, RM(32,16) code have $d_{min}$=8, $N_{min}$=620, the second least distance of 12. q=5, $\{I_1, I_2, I_3, I_4, I_5\} = \{8,12,14,15,16\}$. TABLE III shows the $w_j$ on position (2,N/2+2) and TABLE IV shows the number of codeword with $d_{min}$ for different $T$. It is shown that $T$ reduces the number of codewords with weight of 8 from 620 to 492.

**Theorem 3**: If $U \times H_N$ have the minimum distance $d_{min}$, and the second least minimum distance larger than $d_{min} + 2p$, then there is a $T$ such that $N_{min}(U \times T \times H_N) \leq N_{min}(U \times H_N)$, where $T$ is an upper-triangular matrix with elements: $T_{i,j} = 0$, if $i > j$; $T_{i,j} = 1$, if $i = j$; $T_{i,j} \in \{0,1\}$, if $i < j$.

*Proof:* Let $\{I_1, I_2, ..., I_K\}$ be information bit indices, and let $D_k$ be all codewords with weight $d_{min}$, where $1 \leq k \leq N_{min}$. Suppose that there are $t_m$ frozen bits with indices $\{F_m^1, F_m^2, ..., F_m^{t_m}\}$ which are larger than $I_m$, i.e., $F_m^i > I_m$, $1 \leq i \leq t_m$. Find all vectors $C = (c_1, c_2, ..., c_{t_m})$ such that $0 < w\left(\sum_{k=1}^{t_m} c_k h_N^{(F_m^k)}\right) \leq p$. For each valid vector C, compute the number of codewords that satisfy $w\left(D_k \oplus \left(\sum_{k=1}^{t} c_k h_N^{(F_k)}\right)\right) > d_{min}$ and also corresponds to information bit at index $I_m$ is "1", i.e., $u_{I_m} = 1$. Select the optimal $C$ which provides the maximum number $w\left(D_k \oplus \left(\sum_{k=1}^{t} c_k h_N^{(F_k)}\right)\right)$, and let this maxima be $w_m(C_{opt}^m)$ among all valid C associated with $I_m$. After computing all of $\{I_1, I_2, ..., I_K\}$, we can obtain $w_1(C_{opt}^1), w_2(C_{opt}^2), ..., w_K(C_{opt}^K)$, a $T$ can be designed to have a number of minimum distances of $N_{min} - max\{w_1(C_{opt}^1), w_2(C_{opt}^2), ..., w_K(C_{opt}^K)\}$. The proof is the same as that of Theorem 2.

## V. CONCLUSIONS AND COMMENTS

In this paper, we prove that any pre-transformation with an upper-triangular matrix (including convolution matrix) does not reduce the code minimum distance. We also prove by a design example that a properly designed pre-transformation can reduce the number of codewords with the minimum distance. For these reasons, the CRC/PC/PAC polar codes can outperform the corresponding non-pre-transformed Polar/RM codes under the ML-type decoding. It is still unknown how to optimize the number of minimum distance by designing $T$.

## REFERENCES

[1] E. Arıkan, "Channel polarization: A method for constructing capacity achieving codes for symmetric binary-input memoryless channels," IEEE Trans. Inform. Theory, vol. 55, pp. 3051–3073, July 2009.
[2] I. Tal and A. Vardy, "List decoding of polar codes," available as online as arXiv: 1206.0050v1.
[3] B. Li, H. Shen and D. Tse, "A RM-Polar codes," available as online as arXiv: 1407.5483v1.
[4] K. Niu, K.Chen, "CRC-aided decoding of polar codes," IEEE Communications Letters, 2012.
[5] H. Zhang et al., "Parity-check polar coding for 5G and beyond," in Proc. IEEE Int. Conf. Commun. (ICC), May 2018, pp. 1–7.
[6] E. Arikan, "From Sequential Decoding to Channel Polarization and Back Again," available as online as arXiv: 1908.09594v1.